\documentclass[11pt]{article}

\usepackage{algorithm}
\usepackage{algorithmic}
\usepackage[authoryear]{natbib}

\usepackage{geometry}                
\usepackage{graphicx}
\usepackage{amssymb}
\usepackage{amsmath}
\usepackage{epstopdf}
\usepackage{dsfont}
\usepackage{caption}
\usepackage{subcaption}

\title{Efficient Randomized Quasi-Monte Carlo Methods For Portfolio Market Risk}

\begin{document}

\maketitle

\begin{center}

\large Halis Sak \footnote[1]{Corresponding author. Tel: +86.512.88161000-4886 \\ 
\textit{Email addresses:} halis.sak@gmail.com (Halis Sak), ismailbsgl@gmail.com (\.{I}smail Ba\c{s}o\u{g}lu)}\\ \normalsize
Department of Mathematical Sciences, Xi'an Jiaotong-Liverpool University, Suzhou, China
\\[6pt]

\large \.{I}smail Ba\c{s}o\u{g}lu\\ \normalsize
School of Economics and Administrative Sciences, \.{I}stanbul Kemerburgaz University, \.{I}stanbul, Turkey\\[6pt]

\end{center}

\begin{abstract}
We consider the problem of simulating loss probabilities and conditional excesses for linear asset portfolios under the $t$-copula model. Although in the literature on market risk management there are papers proposing efficient variance reduction methods for Monte Carlo simulation of portfolio market risk, there is no paper discussing combining the randomized quasi-Monte Carlo method with variance reduction techniques. In this paper, we combine the randomized quasi-Monte Carlo method with importance sampling and stratified importance sampling. Numerical results for realistic portfolio examples suggest that replacing pseudorandom numbers (Monte Carlo) with quasi-random sequences (quasi-Monte Carlo) in the simulations increases the robustness of the estimates once we reduce the effective dimension and the non-smoothness of the integrands. 

Keywords: risk management; quasi-Monte Carlo; importance sampling; stratified sampling; t-copula
\end{abstract}

\section{Introduction}

Market risk management deals with the estimation of loss distribution of a portfolio of assets over a fixed time horizon. The widely used risk measures Value-at-Risk (VaR) and expected shortfall require accurate estimates of loss probability and conditional excess under a realistic model that captures dependence structure of the log-returns of multiple assets. As a flexible and accurate model for the logarithmic returns of stocks, we use the $t$-copula dependence structure and marginals following the generalized hyperbolic distribution \citep[see][]{Embrechts:McNeil:Straumann;2002a,Mashal:Naldi:Zeevi;2003a,Prause;1997a,Glasserman;Heidelberger;Shahabuddin:2002a}.

As there are no closed-form analytical results for loss probability and conditional excess under the $t$-copula model, we need to use a computational method like Monte Carlo simulation. In most cases, Monte Carlo simulation  is a better alternative compared to other methods as it leads to error bounds on the estimated values. Due to the fact that Monte Carlo simulation has a slow convergence rate of $O(1/\sqrt{n})$, we need to increase the efficiency of the estimates using variance reduction techniques. There are papers proposing variance reduction methods in portfolio market risk estimation \citep[see, e.g.,][]{Glasserman;Heidelberger;Shahabuddin:2002a,Broadieetal:2011,Basogluetal;2013}.

An alternative to Monte Carlo simulation is the quasi-Monte Carlo method (QMC), which uses low-discrepancy sequences instead of pseudorandom numbers. The rate of convergence of the quasi-Monte Carlo method is close to $O(1/n)$, which is faster than $O(1/\sqrt{n})$. However, an error bound under plain QMC can not be estimated as low-discrepancy sequences do not have an i.i.d. property. Randomized quasi-Monte Carlo solves this problem by applying a randomization on low-discrepancy sequences.

Randomized quasi-Monte Carlo (RQMC) has been used in pricing extensively \citep[see, e.g.,][]{Birge:1995,Boyleetal:1997}. However, the application of RQMC to measure portfolio risk is rarely found in the literature \citep[see][]{Kreininetal:1998a, JinZhang:2006}. This can be explained by the fact that the integrand in risk management applications is a non-smooth function (e.g., indicator function) of high-dimensional random inputs. (As pointed out by \cite{MorokoffCaflisch:1995}, the performance of quasi-Monte Carlo method diminishes when the integrands are not smooth and high-dimensional.) To compute VaR using QMC, \cite{Kreininetal:1998a} apply principal component analysis to reduce the dimensionality of the risk factor space. \cite{JinZhang:2006} efficiently simulate VaR by smoothing the expectation of an indicator function via  Fourier transformation and then applying RQMC.

The motivation of this paper is to investigate whether RQMC and variance reduction techniques can be combined efficiently for simulating loss probability and conditional excess under the $t$-copula model. In order to solve the problem of high-dimensionality of the integrands, we apply a linear transformation on the random input to reduce the effective dimension. Furthermore, the discontinuity of the simulation integrand is reduced using importance sampling. We finally apply stratification to further improve the accuracy of the estimates. Numerical experiments illustrate the effectiveness of RQMC implementations of variance reduction methods over their Monte Carlo implementations. Although, the methodology of the paper is explained on market risk management under the $t$-copula model, it is much more generally applicable to other fields like credit risk, insurance, and operational risk where $t$-copula models are widely used.

The rest of the paper is organized as follows. Section~2 describes the $t$-copula model for portfolio market risk. Section~3 presents background on efficient Monte Carlo simulation methods for estimating loss probability and conditional excess. Section~4 combines the RQMC method with importance sampling and stratified importance sampling for estimating loss probability and conditional excess. We present numerical results in Section~5. 

\section{Portfolio Market Risk in the $t$-Copula Model}
\label{sec:model}

The essence of any model of portfolio market risk is its ability to capture dependence among assets. In this section, we describe the widely used $t$-copula model \citep[see, e.g., ][]{Glasserman;Heidelberger;Shahabuddin:2002a, Sak;Hormann;Leydold:2010a}.

We are interested in the distribution of losses caused by depreciation of stocks over a fixed time period. The following notation is used in order to represent this distribution.
\begin{itemize}
	\item $D$ = the number of stocks in portfolio
	\item $w_d$ = the weight of the $d$th stock
	\item $X_d$ = the log-return of the $d$th stock
	\item $L = 1- \sum\nolimits_{d=1}^D w_d e^{X_d}$ = portfolio loss (initial value of portfolio is assumed to be equal to one)
\end{itemize}

We assume that we are given a portfolio of stocks with known weights $(w_1,\ldots,w_D)'$ and unknown future log-returns $(X_1,\ldots,X_D)'$. The main objective is to estimate  loss probability $P(L > \tau)$, and conditional excess $E\left[ {L|L > \tau } \right]$, especially at large values of $\tau$.

To model dependence among stocks, we need to introduce dependence among the log-returns. The log-return vector $(X_1,\ldots,X_D)'$ of the stocks is assumed to follow a $t$-copula with $\nu$ degrees of freedom. The dependence is introduced through a multivariate $t$-vector $\mathbf{T} =(T_1,\ldots,T_D)'$  with $\nu$ degrees of freedom. Each log-return is represented as
\begin{equation}
{X_d} = {c_d}G_d^{ - 1}\left( {{F_\nu }\left( {{T_d}} \right)} \right), \, d=1,\ldots,D,
\label{eq:logreturns}
\end{equation}
in which
\begin{itemize}
\item $F_{\nu}$ denotes the cumulative distribution function (CDF) of a $t$-distribution with $\nu$ degrees of freedom;
\item $G_d$ denotes the CDF of the marginal distribution of the $d$th log-return; 
\item $c_d$ is the scaling factor for the $d$th log-return.
\end{itemize}

Through this representation, the dependence among the log-returns, $X_d$, can be determined by the correlations among $T_d$. Suppose, we are given the correlation matrix $\mathbf{\Sigma}$ of vector $\mathbf{T}$ and let $\mathbf{\Lambda} \in \mathds{R}^{D\times D}$ be the lower triangular Cholesky factor of $\mathbf{\Sigma}$ satisfying $\mathbf{\Lambda}\mathbf{\Lambda}'=\mathbf{\Sigma}$. Then, $\mathbf{T}$ can be generated using
\begin{equation}
\mathbf T = \frac{{\mathbf{\Lambda Z}}}{{\sqrt {{Y \mathord{\left/
 {\vphantom {Y \nu }} \right.
 \kern-\nulldelimiterspace} \nu }} }},
\label{eq:multit}
\end{equation}
where $\mathbf{Z}=(Z_1,\ldots,Z_D)'$ is a standard multi-normal random vector and $Y$ is an independent chi-squared random variable  with $\nu$ degrees of freedom.

\section{Efficient Monte Carlo Simulation Methods}
\label{sec:effmc}

In this section, we provide a brief summary of efficient Monte Carlo simulation algorithms designed for the estimation of portfolio market risk. Before that, we start with the implementation of the naive Monte Carlo algorithm. The naive identity of the loss probability is $P\left( {L > \tau } \right) = E\left[ {{\mathbf 1_{\left\{ {L > \tau } \right\}}}} \right]$, where $\mathbf 1\left\{.\right\}$ denotes the indicator of the event in braces. We are also interested in the conditional excess that can be represented as the ratio of two expectations
\begin{equation}
E\left[ {L|L > \tau } \right] = \frac{{E\left[ {L{\mathbf 1_{\left\{ {L > \tau } \right\}}}} \right]}}{{P\left( {L > \tau } \right)}} = \frac{{E\left[ {L{\mathbf 1_{\left\{ {L > \tau } \right\}}}} \right]}}{{E\left[ {{\mathbf 1_{\left\{ {L > \tau } \right\}}}} \right]}},
\label{eq:ratioest}
\end{equation}
which can be estimated in a single simulation run.

Each replication of the naive Monte Carlo algorithm follows the steps given below:
\begin{enumerate}
	\item[1.] Generate $D$ independent standard normal random variables, $\mathbf Z =\left(Z_1,\ldots,Z_D\right)'$, and a chi-squared random variable $Y$ with $\nu$ degrees of freedom, independent of $\mathbf Z$.
	\item[2.] Calculate $\mathbf T$ in (\ref{eq:multit}).
	\item[3.] Calculate the log-returns $X_d$, $d=1,\ldots,D$ in (\ref{eq:logreturns}).
	\item[4.] Compute the portfolio loss $L = 1- \sum\nolimits_{d=1}^D w_d \exp\left(X_d\right)$ and return the estimators ${\mathbf 1_{\left\{ {L > \tau } \right\}}}$ and $L{\mathbf 1_{\left\{ {L > \tau } \right\}}}$. 
\end{enumerate}
 
\subsection{Importance Sampling}

At a large threshold value $\tau$, most of the replications of the naive simulation algorithm return the value zero for the estimator ${\mathbf 1_{\left\{ {L > \tau } \right\}}}$. To increase the number of replications that fall in the region $L>\tau$, importance sampling modifies the joint density of the random input.

Suppose $f\left(.\right)$ is the joint probability density function (PDF) of input variables $\mathbf Z$ and $Y$, and $\tilde f\left(.\right)$ is the modified density. Importance sampling uses the following identity to estimate the loss probability
\[E\left[ {{{\bf{1}}_{\left\{ {L > \tau } \right\}}}} \right] = \tilde E\left[ {{{\bf{1}}_{\left\{ {L > \tau } \right\}}}\frac{{f\left( {{\bf{Z}},Y} \right)}}{{\tilde f\left( {{\bf{Z}},Y} \right)}}} \right],\]
where $\tilde E$ is the expectation taken using the modified density $\tilde f\left(.\right)$.

Finding an importance sampling density that minimizes the variance of Monte Carlo estimators is a subtle problem. But it is possible to use the zero-variance IS function in search of an effective IS density (see, e.g., \cite{Glasserman;Heidelberger;Shahabuddin:1999a} and \cite{Arouna:2004}). \cite{Glasserman;Heidelberger;Shahabuddin:1999a} add the mode of the zero-variance IS function as a mean shift to the original density for pricing path-dependent options. \cite{Sak;Hormann;Leydold:2010a} utilize the same idea to find a close-to-optimal optimal parameters for simulating loss probabilities in the $t$-copula model of portfolio market risk. 

\cite{Sak;Hormann;Leydold:2010a} add a mean shift vector with negative
entries to the normal vector $\mathbf Z$ and use a scale parameter less than two for the chi-square (i.e., Gamma) random variable $Y$ to construct the IS density. The shift vector and the scale value are selected so that the mode of the resulting
IS density is equal to the mode of the zero-variance IS function. For more details on the determination of the IS parameters and implementation of the simulation algorithm, see Section~4 of \cite{Sak;Hormann;Leydold:2010a}.

\subsection{Stratified Importance Sampling}
\label{sec:sis}

To obtain further variance reduction, one can stratify the importance sampling density along one or possibly more directions. For the $t$-copula model, suppose that $\xi_i$, $i = 1, \ldots , I$, is a partition of $\mathds R^{D+1}$ into $I$ disjoint subsets with probabilities $\tilde p_i=\tilde P\left(\left(\mathbf Z,Y\right) \in \xi_i\right)$ under the IS density. Then the stratified importance sampling (SIS) identity is given by
\[\tilde E\left[ {{{\bf{1}}_{\left\{ {L > \tau } \right\}}}\frac{{f\left( {{\bf{Z}},Y} \right)}}{{\tilde f\left( {{\bf{Z}},Y} \right)}}} \right] = \sum\limits_{i = 1}^I {{{\tilde p}_i}\tilde E\left[ {{{\bf{1}}_{\left\{ {L > \tau } \right\}}}\frac{{f\left( {{\bf{Z}},Y} \right)}}{{\tilde f\left( {{\bf{Z}},Y} \right)}}|\left( {{\bf{Z}},Y} \right) \in {\xi _i}} \right]}. \]

Although stratified sampling is a simple variance reduction technique, its performance is adversely affected by the high-dimensionality of the sample space \citep[see][]{Cheng;Davenport:1989}. Thus, we need to reduce the effective dimension for stratification. Under the $t$-copula setting for portfolio market risk, \cite{Basogluetal;2013} reduce the number of stratified dimensions from $D+1$ to two by stratifying $\mathbf Z$ along a single direction and stratifying $Y$. They use the IS shift of \cite{Sak;Hormann;Leydold:2010a} as the stratification direction for $\mathbf Z$, since the IS shift provides a good stratification direction for multivariate normal input \citep[see][]{Glasserman;Heidelberger;Shahabuddin:1999a}.

To elaborate on the implementation of stratification in \cite{Basogluetal;2013}, they use equiprobable strata and minimize the variance of the stratified estimator using an optimal sample allocation. For equiprobable strata, the optimal allocation of replications to a stratum is proportional to the conditional standard deviation of that stratum \citep[see, e.g.,][page 217]{Glasserman:2004a}. Estimates of standard deviations can be computed using a pilot run. An iterative version of the same idea is presented in \cite{Etore:Jourdain;2010a} under the name of adaptive optimal allocation (AOA). In each iteration, the AOA algorithm modifies the proportion of further replications by using conditional standard deviation estimates. These proportions converge to the optimal allocation fractions through the iterations. \cite{Etore:Jourdain;2010a} show that the stratified estimator of the AOA algorithm is asymptotically normal and its asymptotic variance is minimal. \cite{Basogluetal;2013} utilize the AOA algorithm of \cite{Etore:Jourdain;2010a} with minor modifications. For more details on the estimation of loss probabilities using the stratified importance sampling method, see Section~5 of \cite{Basogluetal;2013}.

For conditional excess simulation, the allocation strategy of the SIS algorithm should be different than the one that we use for loss probability simulation. In that case, we use optimal allocation fractions that minimize the variance of the stratified ratio estimator of conditional excess \citep[see][]{BasogluHormann:2014}. For equiprobable strata, the allocation fractions should be proportional to
\[\frac{{{x^2}s_{i,y}^2}}{{{y^4}}} - \frac{{2x{s_{i,xy}}}}{{{y^3}}} + \frac{{s_{i,x}^2}}{{{y^2}}},\,i = 1, \ldots ,I,\]
where $x=E\left[ {{L\mathbf 1_{\left\{ {L > \tau } \right\}}}} \right]$, $y=E\left[ {{\mathbf 1_{\left\{ {L > \tau } \right\}}}} \right]$, ${s_{i,x}^2}$ and ${s_{i,y}^2}$ are the variances of ${{L\mathbf 1_{\left\{ {L > \tau } \right\}}}}$ and ${{\mathbf 1_{\left\{ {L > \tau } \right\}}}}$ conditional on the $i$th stratum, and ${s_{i,xy}}$ is the covariance of ${{L\mathbf 1_{\left\{ {L > \tau } \right\}}}}$ and ${{\mathbf 1_{\left\{ {L > \tau } \right\}}}}$ conditional on the $i$th stratum. These values can be estimated through the iterations of the AOA algorithm.

\section{Improving The Efficiency Using RQMC}

In this section, we first shortly describe the difference between randomized quasi-Monte Carlo and Monte Carlo simulation adapted to the problem of estimating portfolio market risk. In Monte Carlo simulation, we randomly sample points from $\left[0, 1\right)^{D+1}$ to approximate the integrals (this is implicitly done while generating $Y$ and $\mathbf Z$). Quasi-Monte Carlo sampling utilizes sample points in $\left[0, 1\right)^{D+1}$ that come from a low-discrepancy point set. In contrast to a Monte Carlo sample, low-discrepancy point sets do not have the i.i.d. property. Therefore, we cannot directly employ the error bound formula used in Monte Carlo simulation. However, a random sample of quasi-random estimators can be constructed based on a low-discrepancy point set. This can be achieved by creating independent copies of a low-discrepancy point set $P_N =\left\{\mathbf U_1,\ldots,\mathbf U_{N}\right\}$ by the following randomization

\begin{equation}
{\tilde{\mathbf{U}}_i} = \left( {{{\mathbf{U}}_i} + {\bf{W}}} \right)\bmod 1,
\label{eq:shift}
\end{equation}
where $\mathbf W$ is a uniformly distributed vector in $\left[0, 1\right)^{D+1}$. 

The vector $\tilde{\mathbf{U}}_i$ is uniformly distributed in the unit hypercube \citep[see][page 204]{Lemieux:2009}. Thus, the estimators based on $\tilde{\mathbf{U}}_i$ are unbiased, and the error bounds for the estimates can be obtained using $M$ independent randomized copies of $P_N$.

Given a low-discrepancy point sequence $P_N$ in $\left[0, 1\right)^{D+1}$, each replication of the randomized quasi-Monte Carlo version of the naive simulation algorithm (see Section~\ref{sec:effmc}) follows the steps given below:
\begin{enumerate}
  \item[1.] Generate a randomized copy $\tilde P_{N} = \left\{ \mathbf {\tilde U}_1,\ldots,\mathbf {\tilde U}_{N} \right\}$ of $P_N=\left\{\mathbf U_1,\ldots,\mathbf U_{N}\right\}$ using (\ref{eq:shift}).
	\item[2.] For $i=1,\ldots,N$;
	\begin{enumerate}
    \item[a.] Compute ${Y^{\left(i\right)}} = F_\Gamma ^{ - 1}\left( {{{\tilde U}_{i,1}};\frac{\nu }{2},2} \right)$ using the inverse CDF $F_\Gamma ^{ - 1}$ of the Gamma distribution.
	  \item[b.] Compute ${Z^{\left(i\right)}_d} = \Phi^{ - 1}\left( {{{\tilde U}_{i,d+1}}} \right)$ for $d=1,\ldots,D$, using the inverse CDF $\Phi^{ - 1}$ of the standard normal distribution.
	  \item[c.] Calculate $\mathbf T^{\left(i\right)}$ in (\ref{eq:multit}).
	  \item[d.] Calculate the log-returns $X^{\left(i\right)}_d$, $d=1,\ldots,D$, in (\ref{eq:logreturns}).
	  \item[e.] Compute the portfolio loss $L^{\left(i\right)} = 1- \sum\nolimits_{d=1}^D w_d \exp\left(X^{\left(i\right)}_d\right)$. 
	\end{enumerate}
  \item[3.] Return the estimators $N^{-1}\sum\nolimits_{i=1}^{N}{\mathbf 1_{\left\{ {L^{\left(i\right)} > \tau } \right\}}}$ and $N^{-1}\sum\nolimits_{i=1}^{N}{L^{\left(i\right)}{\mathbf 1_{\left\{ {L^{\left(i\right)} > \tau } \right\}}}}$. 
\end{enumerate}

As pointed out by \cite{Caflisch:1998}, the performance of QMC methods is degraded by the high dimension and the non-smoothness of the integrand. The RQMC implementation given above suffer from these problems since the number of stocks $D$ in a portfolio can be large and the integrands contain an indicator function. In order to improve the efficiency of RQMC estimators, one needs to solve these problems or decrease their impact.

The first problem, high-dimensionality, can be solved by applying a linear transformation to the vector $\mathbf Z$ so that the effective dimension of the integrand can be reduced \citep[see][]{Imai;Tan:2014}. Furthermore, the impact of the second problem, non-smoothness, can be reduced using variance reduction techniques \citep[see, e.g.][for an application in Asian option pricing]{DingecHormann:2014}. In the rest of the section, we first explain the linear transformation methodology that we use. Then, we describe how RQMC can be efficiently combined with IS and SIS.

\subsection{Linear Transformation}

The natural implementation of RQMC on naive simulation given above simply changes pseudorandom numbers with randomized low-discrepancy point sets. Our numerical experiments indicate that this approach does not yield a significant improvement over the naive Monte Carlo method. This is due to the high dimension of the problem, i.e., every random input ($Z_1,\ldots,Z_D,Y$) has a significant contribution to the variance of the estimators. In order to increase the efficiency of RQMC on IS, the effective dimension of the problem should be reduced such that the most of the variance can be explained by only a few random inputs. Then, these random inputs can be generated through the first few elements of the randomized low-discrepancy points. The rationale behind this is the fact that the first low dimensional projections of the low-discrepancy point sets have better uniformity properties \citep[see][]{Caflisch:1998}.

We apply a linear transformation to the random vector $\mathbf Z$ so that the first element $Z_1$ corresponds to the IS shift $\boldsymbol \mu$ given in \cite{Sak;Hormann;Leydold:2010a}. The linear transformation can be applied by first multiplying $\mathbf Z$ with an orthogonal matrix $\mathbf V \in \mathds R^{ D\times D}$ that has its first column equal to $\mathbf v=\boldsymbol \mu/\left\| \boldsymbol \mu  \right\|$. The remaining columns can be arbitrarily selected. This transformation increases the impact of $Z_1$ on the variance of the estimators. Then, we use the first two elements of the randomized low-discrepancy points to generate $Y$ and $Z_1$, respectively.

In the application of linear transformation, the only change in the replications of the naive RQMC algorithm is the computation of the $\mathbf T$ vector using
\begin{equation}
{\bf{T}} = \frac{{\mathbf \Lambda {\bf{VZ}}}}{{\sqrt {{Y \mathord{\left/
 {\vphantom {Y \nu }} \right.
 \kern-\nulldelimiterspace} \nu }} }} = \frac{{{\bf{AZ}}}}{{\sqrt {{Y \mathord{\left/
 {\vphantom {Y \nu }} \right.
 \kern-\nulldelimiterspace} \nu }} }},
\label{eq:lt}
\end{equation}
where $\mathbf A= \mathbf \Lambda \mathbf V$. 

\subsection{RQMC on Importance Sampling}
\label{sec:rqmcis}

The linear transformation described above reduces the effective dimension of the integrand considerably. But, as it was mentioned earlier, the non-smoothness of the integrands, caused by the indicator function ${\mathbf 1_{\left\{ {L > \tau } \right\}}}$, still has a strong impact on the performance of RQMC. At a large threshold value $\tau$, most of the replications of the naive simulation algorithm return zero values for the estimator ${\mathbf 1_{\left\{ {L > \tau } \right\}}}$. This results in large-sized jumps in the naive Monte Carlo integrands.

\begin{figure}[htbp]
    \centering
    \begin{subfigure}[b]{0.48\textwidth}
        \includegraphics[width=\textwidth]{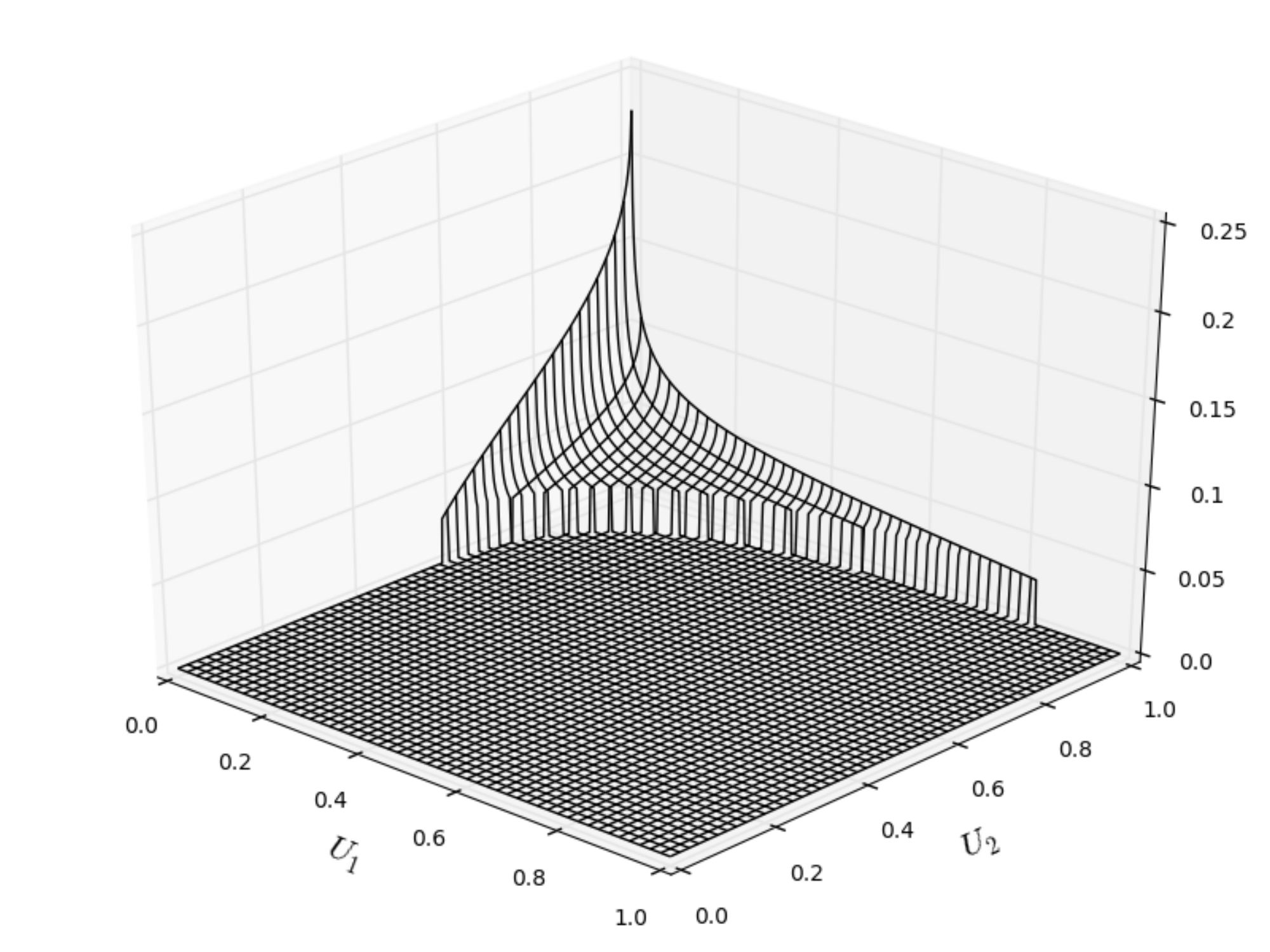}
        \caption{Naive Monte Carlo Integrand}
        \label{fig:2DEXC_NV}
    \end{subfigure}
    ~ 
    \begin{subfigure}[b]{0.48\textwidth}
        \includegraphics[width=\textwidth]{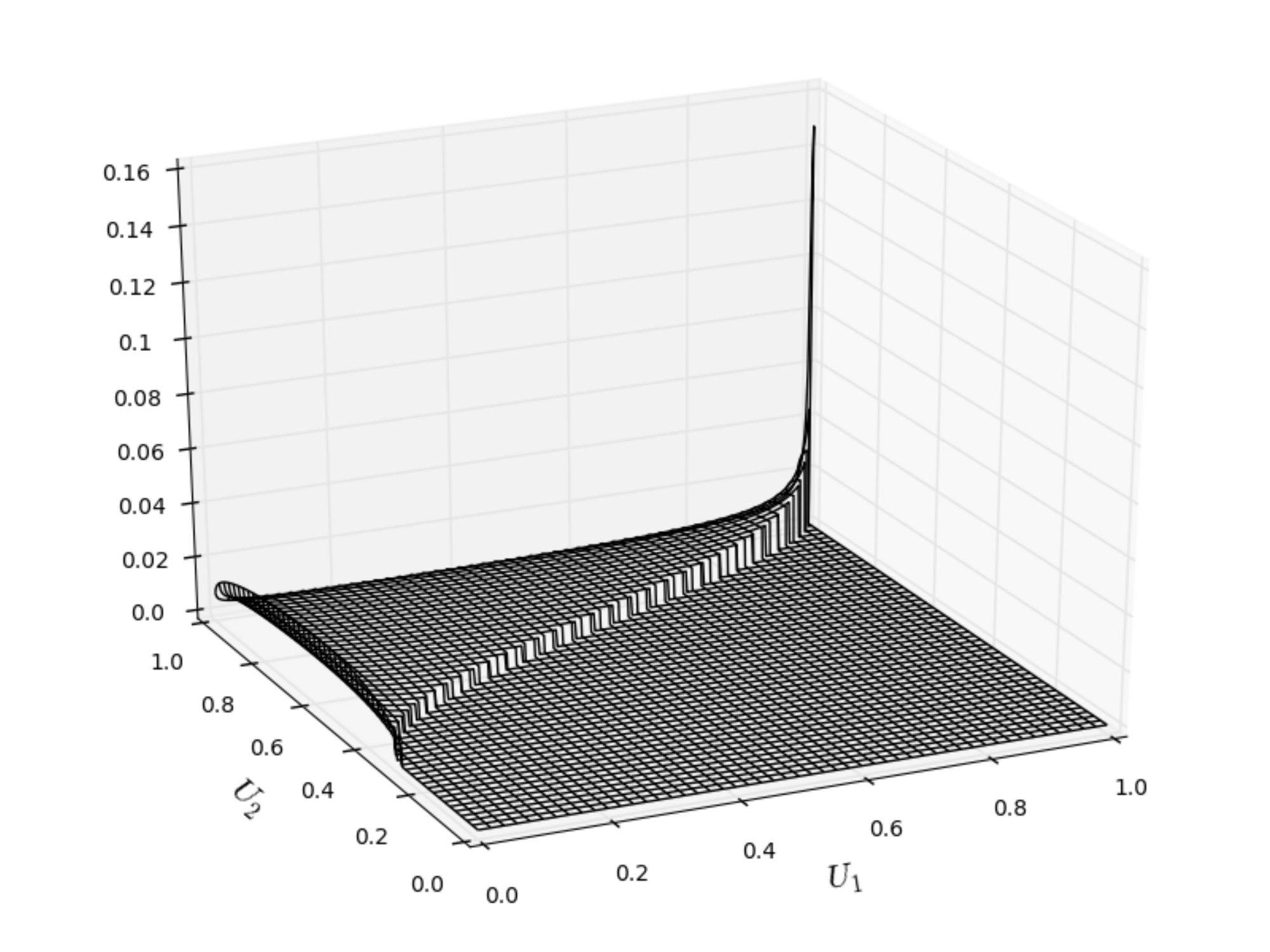}
        \caption{IS Integrand}
        \label{fig:2DEXC_IS}
    \end{subfigure}
    \caption{Monte Carlo integrands of $L\mathbf 1_{\left\{L>\tau\right\}}$ on $\left[0,1\right)^2$.}
		\label{fig:integrand}
\end{figure}

Figure~\ref{fig:2DEXC_NV} illustrates the non-smoothness problem of the naive integrand $L\mathbf 1_{\left\{L>\tau\right\}}$ for a numerical example of Section~\ref{sec:numres}. We use the portfolio that consists of two stocks with $t$ marginals (for the parameter values, see Section~\ref{sec:numres}). In this case, the integrand $L\mathbf 1_{\left\{L>\tau\right\}}$ is a function on $\left[0,1\right)^3$. Indeed, such an integrand can not be illustrated in three-dimensional space. However, for demonstration purposes, it is possible to fix $Z_2$ to zero ($U_3=0.5$) as the impact of $Z_2$ on the integrand is considerably reduced after the linear transformation.

Figure~\ref{fig:2DEXC_IS} illustrates the Monte Carlo integrand under importance sampling density. The main difference between Figure~\ref{fig:2DEXC_IS} and Figure~\ref{fig:2DEXC_NV} is the reduced size of the jumps for most of the domain. We achieve this by multiplying the integrand $L\mathbf 1_{\left\{L>\tau\right\}}$ with the likelihood ratio $f\left(\mathbf Z, Y\right)/{\tilde f}\left(\mathbf Z,Y\right)$ in IS.

In the RQMC implementation of IS, we simply shift $Z_1$ with $\left\| \boldsymbol \mu  \right\|$ and generate $Y$ under the IS scale parameter $\theta$. Then, we apply the linear transform to $\mathbf Z$ using (\ref{eq:lt}). Finally, the responses ${\mathbf 1_{\left\{ {L > \tau } \right\}}}$ and $L{\mathbf 1_{\left\{ {L > \tau } \right\}}}$ are multiplied with the likelihood ratio
\[\frac{{f\left( {{\bf{Z}},Y} \right)}}{{\tilde f\left( {{\bf{Z}},Y} \right)}} = \exp \left( {\frac{1}{2}{{\left\| \boldsymbol \mu  \right\|}^2} - {Z_1}\left\| \boldsymbol \mu  \right\| + \frac{{\left( {2 - \theta } \right)Y}}{{2\theta }} + \frac{\nu }{2}\log \left( {\frac{\theta }{2}} \right)} \right).\]

\subsection{RQMC on Stratified Importance Sampling}
\label{sec:rqmcsis}

As observed in Figure~\ref{fig:2DEXC_IS}, importance sampling reduces the size of the jumps over most of the integrand domain. However, if we focus on the region where both $U_1$ and $U_2$ take values close to one, we see large-sized jumps. The variance contribution of such regions on the IS estimator is significant. To remedy this problem, one may allocate more replications to those regions through stratification. In previous subsections, we have explained how RQMC can be combined with linear transformation and IS. This subsection describes how stratification can be combined with the previously discussed techniques.

In the RQMC version of the SIS algorithm, we utilize the same low-discrepancy point sequence in each stratum. To guarantee independence across strata, we use different random shifts for each stratum. These random shifts remain the same throughout the iterations of the AOA algorithm. When the algorithm decides to allocate more replications in a stratum, we start with the first point of the low-discrepancy sequence that has not been used in the previous iterations. Furthermore, the sample allocation decision is made based on the optimal allocation fractions described in Section~\ref{sec:sis}.

The error bound for the randomized quasi-Monte Carlo SIS estimator can be computed using the $M$ outer replications of the algorithm.

\section{Numerical Results}
\label{sec:numres}

In order to illustrate the efficiency of the randomized quasi-Monte Carlo method, we implement all algorithms in R \citep{rcoreteam}. To generate randomized low-discrepancy point sets, we employ randomly shifted Sobol nets using the implementation of \cite{Bratley:Fox;1988} using the R-package ``randtoolbox'' \citep{randtoolbox}.

In our experiments, we use stock portfolios with sizes $D$ equal to 2, 5, and 10. For the choice of marginal distributions, we use the generalized hyperbolic and the $t$ distribution, as they seem to be the best fitting distributions for the stock log-returns. For the model parameters, we use the fitted values for NYSE data reported in \cite{Halulu:2012a} (see page 65 for the list of stocks, Tables E.1, E.6, E.7, and E.8 for the $t$-copula parameters, and Tables 6.4 and 6.5 for the marginal parameters). 

We present $95\%$ error bounds of naive ($EB_{NV}$), IS ($EB_{IS}$), and SIS ($EB_{SIS}$) MC simulations and naive ($EB_{QNV}$), linear transformation ($EB_{QLT}$), and IS ($EB_{QIS}$) RQMC simulations for loss probability and conditional excess estimation using the $t$ and the generalized hyperbolic (GH) marginals in Table~1. Threshold values ($\tau$) that gives loss probabilities of $0.05$ and $0.001$ are also provided for different parameter settings. For each parameter setting, the first row gives the error bounds for loss probability estimates and the second row for the conditional excess estimates. Note that the loss probability and the conditional excess simulations are performed separately.

In these experiments, the total number of replications used is $n=10^5$ for naive and IS simulations and approximately $n\approx 10^5$ for SIS simulation. We terminate SIS in four iterations, using approximately 10, 20, 30, and 40 percent of the total sample size in each iteration, sequentially. The number of outer replications in RQMC simulations is selected as $M=40$ in order to give reliable error bounds on the estimate. Thus, we do $N=n/M=2,500$ inner replications. For this setting, the randomized quasi-Monte Carlo version of stratified importance sampling does not provide reliable error bounds, since $N=2,500$ is not sufficiently large to satisfy the asymptotic normality of the stratified estimators. For this reason, we do not provide error bounds of the randomized quasi-Monte Carlo version of stratified importance sampling. In the following paragraphs, we provide information on the convergence properties of the randomized quasi-Monte Carlo version of stratified importance sampling.

\begin{table}[htbp]
	\centering
		\resizebox{15cm}{!}{
		\begin{tabular}{l*{18}{c}}
		\hline
		& & & \multicolumn{7}{c}{Loss probability $\approx$ 0.05} & &  \multicolumn{7}{c}{Loss probability $\approx$ 0.001} \\ 
		\cline{4-10} \cline{12-18}
		Marginals &  $D$&&$\tau$&$EB_{NV}$&$EB_{IS}$&$EB_{SIS}$&$EB_{QNV}$&$EB_{QLT}$&$EB_{QIS}$&&$\tau$&$EB_{NV}$&$EB_{IS}$&$EB_{SIS}$&$EB_{QNV}$&$EB_{QLT}$&$EB_{QIS}$\\ 
		\hline
t	&	2	&	&	0.0271	&	1.4E-03	&	5.8E-04	&	1.3E-04	&	6.0E-04	&	4.3E-04	&	2.5E-04	&		&	0.0942	&	2.1E-04	&	1.3E-05	&	3.7E-06	&	1.3E-04	&	1.5E-04	&	6.6E-06	\\
	&		&	&		&	5.0E-04	&	1.8E-04	&	5.9E-05	&	2.7E-04	&	1.9E-04	&	8.4E-05	&		&		&	5.9E-03	&	4.5E-04	&	1.6E-04	&	8.5E-03	&	6.8E-03	&	2.1E-04	\\
	&	5	&	&	0.0264	&	1.4E-03	&	4.5E-04	&	1.7E-04	&	7.7E-04	&	4.5E-04	&	2.1E-04	&		&	0.1258	&	2.0E-04	&	1.2E-05	&	5.2E-06	&	1.2E-04	&	1.4E-04	&	8.8E-06	\\
	&		&	&		&	8.0E-04	&	2.0E-04	&	8.7E-05	&	3.8E-04	&	3.2E-04	&	8.2E-05	&		&		&	1.5E-02	&	6.3E-04	&	2.9E-04	&	9.3E-03	&	1.0E-02	&	4.5E-04	\\
	&	10	&	&	0.0232	&	1.3E-03	&	4.4E-04	&	1.4E-04	&	7.3E-04	&	4.1E-04	&	1.9E-04	&		&	0.0784	&	1.9E-04	&	1.2E-05	&	4.9E-06	&	1.5E-04	&	1.4E-04	&	7.3E-06	\\
	&		&	&		&	4.0E-04	&	1.1E-04	&	4.2E-05	&	3.1E-04	&	1.6E-04	&	4.2E-05	&		&		&	4.6E-03	&	3.0E-04	&	1.4E-04	&	7.7E-03	&	4.8E-03	&	1.7E-04	\\
		\hline
GH	&	2	&	&	0.0280	&	1.3E-03	&	5.4E-04	&	1.3E-04	&	5.9E-04	&	4.7E-04	&	2.4E-04	&		&	0.0957	&	1.8E-04	&	1.4E-05	&	4.1E-06	&	1.3E-04	&	1.1E-04	&	7.3E-06	\\
	&		&	&		&	4.7E-04	&	1.7E-04	&	5.6E-05	&	2.4E-04	&	2.1E-04	&	8.3E-05	&		&		&	4.3E-03	&	3.5E-04	&	1.3E-04	&	5.5E-03	&	4.2E-03	&	1.8E-04	\\
	&	5	&	&	0.0270	&	1.4E-03	&	4.5E-04	&	1.7E-04	&	6.3E-04	&	4.9E-04	&	2.2E-04	&		&	0.1060	&	2.1E-04	&	1.2E-05	&	4.9E-06	&	1.5E-04	&	1.3E-04	&	6.3E-06	\\
	&		&	&		&	6.1E-04	&	1.6E-04	&	7.2E-05	&	3.2E-04	&	2.4E-04	&	8.0E-05	&		&		&	6.7E-03	&	4.0E-04	&	1.9E-04	&	6.9E-03	&	9.0E-03	&	2.5E-04	\\
	&	10	&	&	0.0235	&	1.4E-03	&	4.4E-04	&	1.4E-04	&	8.1E-04	&	3.4E-04	&	1.7E-04	&		&	0.0689	&	1.9E-04	&	1.2E-05	&	4.0E-06	&	1.5E-04	&	1.5E-04	&	6.3E-06	\\
	&		&	&		&	3.1E-04	&	9.0E-05	&	3.5E-05	&	1.9E-04	&	1.3E-04	&	3.6E-05	&		&		&	3.0E-03	&	1.5E-04	&	6.5E-05	&	3.4E-03	&	3.3E-03	&	8.9E-05	\\
			
		\hline
		\end{tabular}
		}
		\caption{ $95\%$ error bounds of naive ($EB_{NV}$), IS ($EB_{IS}$), and SIS ($EB_{SIS}$) MC simulations and naive ($EB_{QNV}$), linear transformation ($EB_{QLT}$), and IS ($EB_{QIS}$) RQMC simulations for loss probability and conditional excess estimation using the $t$ and the generalized hyperbolic marginals.}
	\label{tab:TLP}
\end{table}

From Table~\ref{tab:TLP}, we observe that the use of RQMC sampling clearly reduces the error bound of the naive MC for loss probability estimates, however, the improvement is less clear for probability of 0.001. For conditional excess simulation, RQMC implementation is only efficient for loss probability of 0.05. Combination of RQMC with linear transformation further reduces the error bounds of the estimates when the loss probability level is 0.05. When we compare the error bounds of Monte Carlo IS with RQMC IS, we observe a reasonable improvement for all cases. The reason behind this performance boost of RQMC over MC is the fact that the IS integrand is smoother (see Section~\ref{sec:rqmcis}).   

\begin{figure}[htbp]
    \centering
    \begin{subfigure}[b]{0.48\textwidth}
        \includegraphics[width=\textwidth]{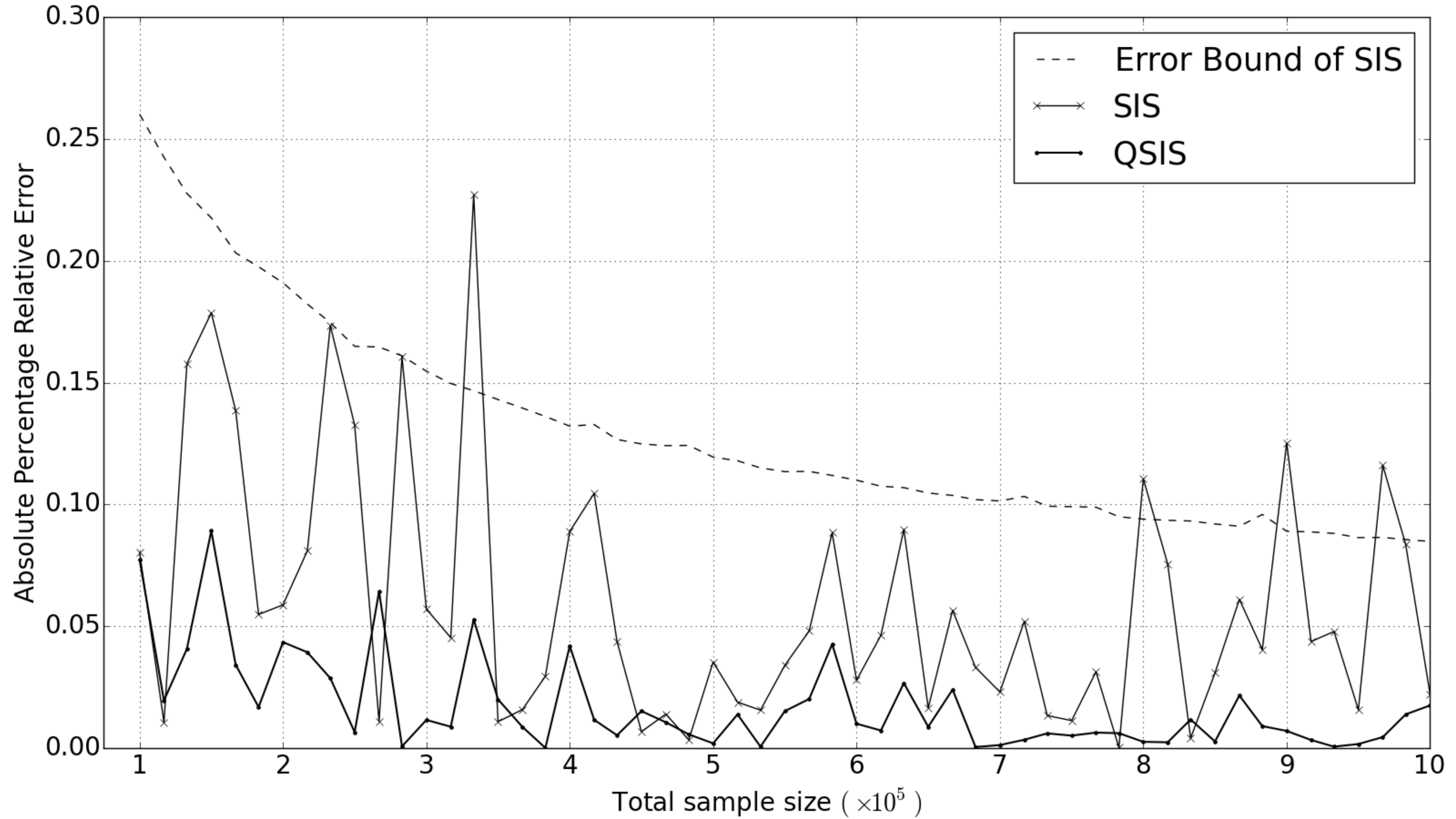}
        \caption{Loss probability simulation ($D=2$)}
        \label{fig:apreD_2_TLP}
    \end{subfigure}
    ~ 
    \begin{subfigure}[b]{0.48\textwidth}
        \includegraphics[width=\textwidth]{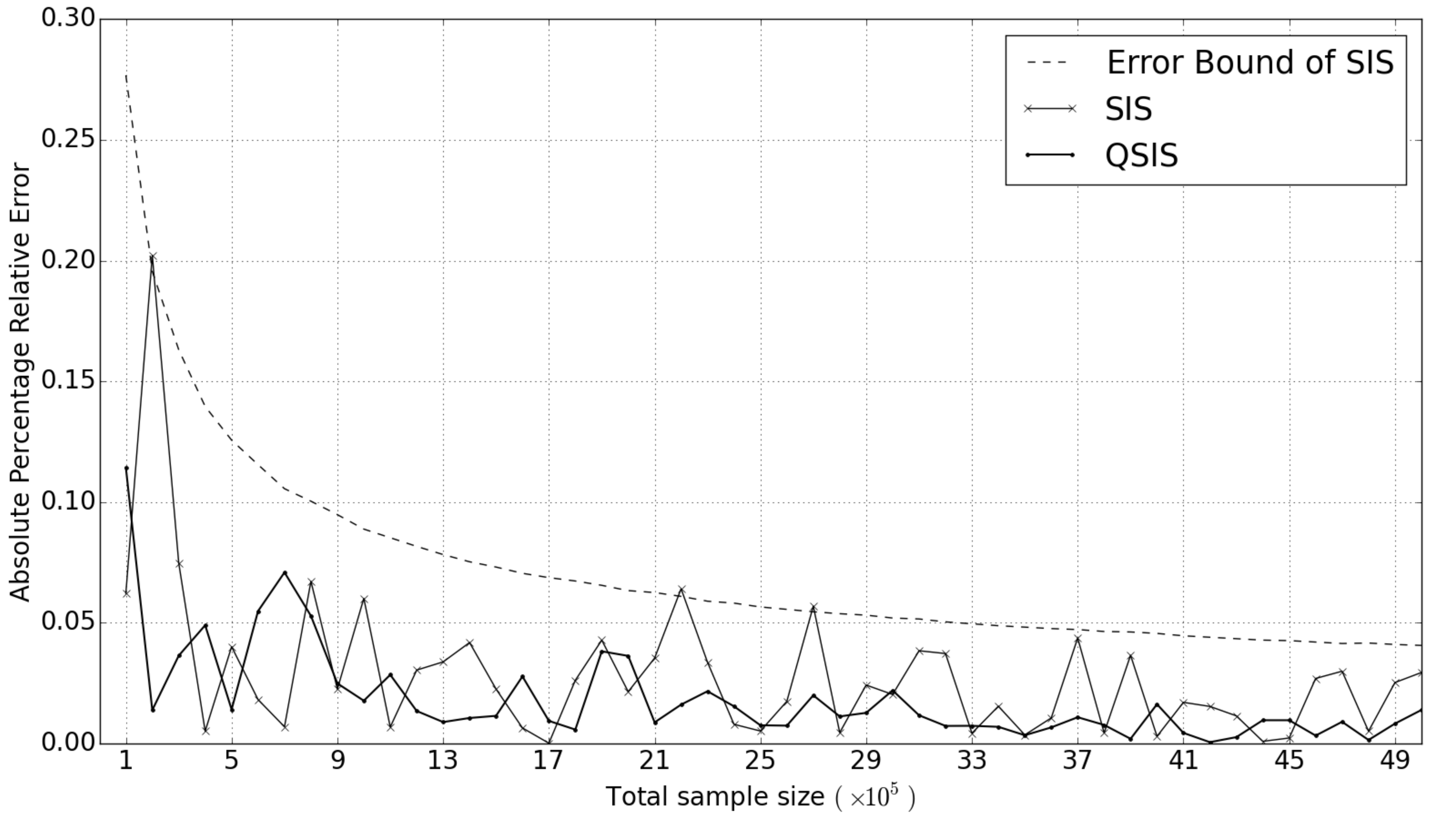}
        \caption{Loss probability simulation ($D=10$)}
        \label{fig:apreD_10_TLP}
    \end{subfigure}
		
    \begin{subfigure}[b]{0.48\textwidth}
        \includegraphics[width=\textwidth]{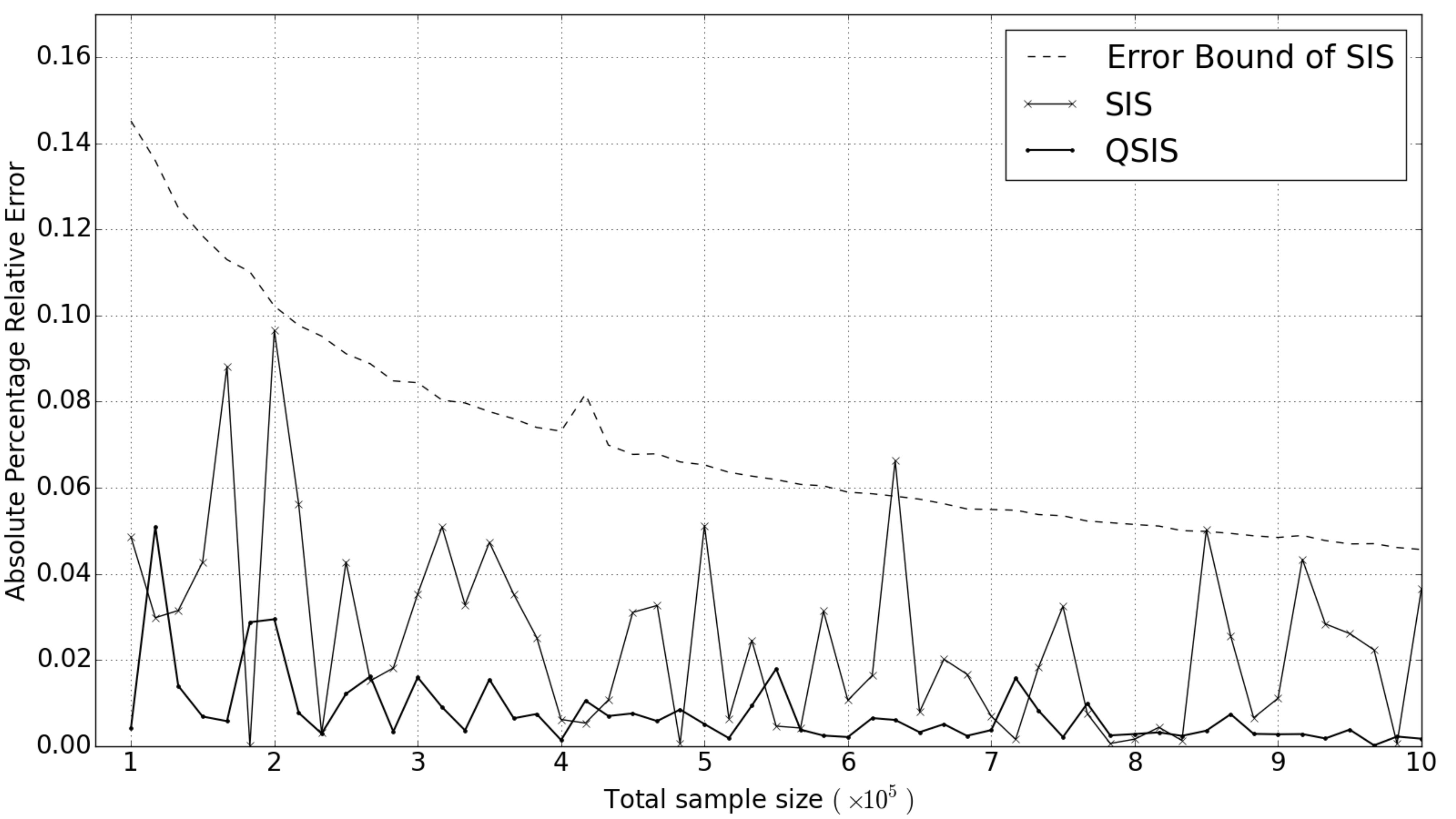}
        \caption{Conditional excess simulation ($D=2$)}
        \label{fig:apreD_2_CE}
    \end{subfigure}
    ~ 
    \begin{subfigure}[b]{0.48\textwidth}
        \includegraphics[width=\textwidth]{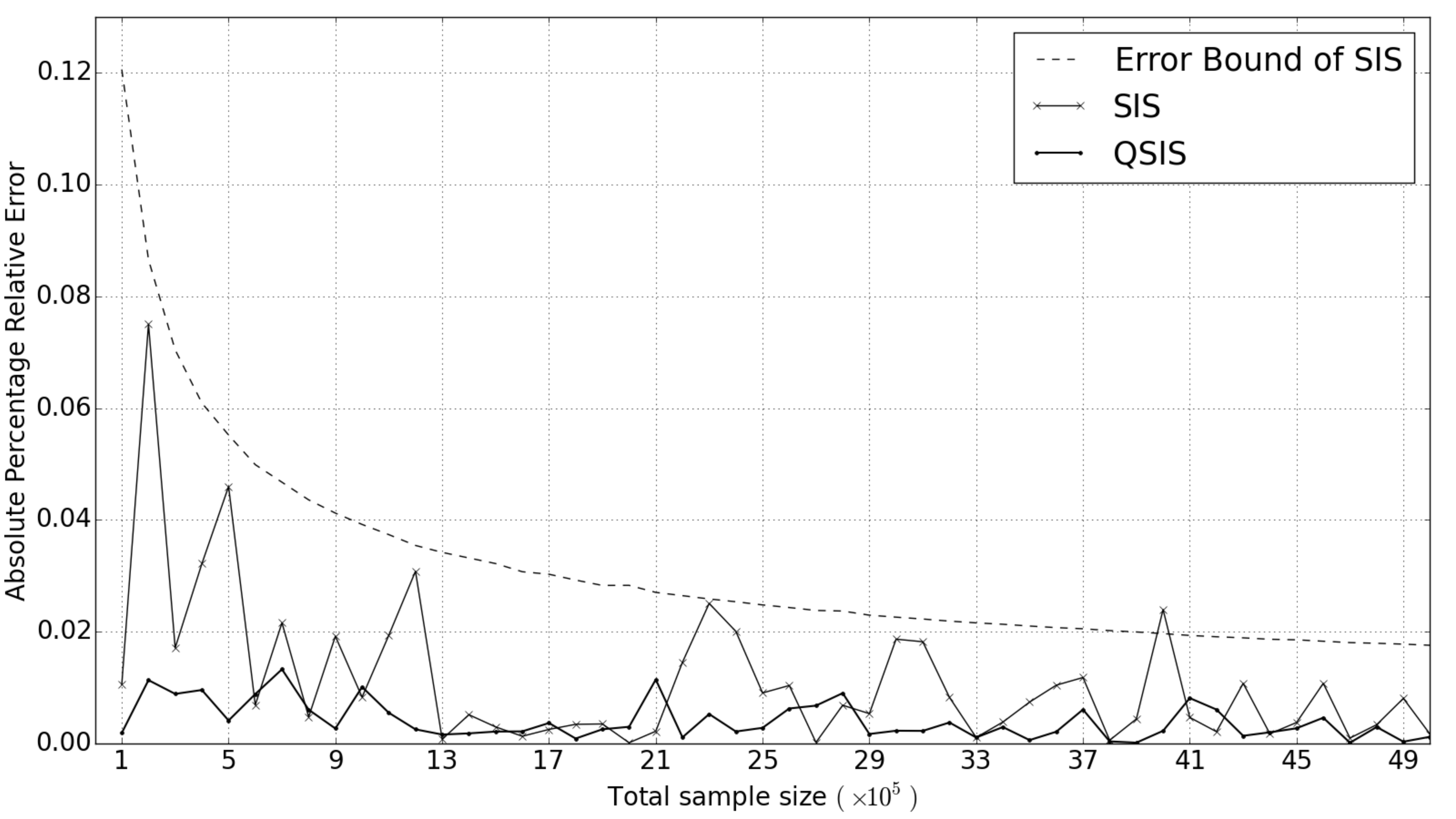}
        \caption{Conditional excess simulation ($D=10$)}
        \label{fig:apreD_10_CE}
    \end{subfigure}
    \caption{The absolute percentage relative errors for MC SIS and RQMC SIS estimators under the $t$ marginals for loss probability of 0.05.}\label{fig:convergence}
\end{figure}

Figure~\ref{fig:convergence} shows the convergence of RQMC SIS estimates to the nearly exact values (we use $n=10^9$ replications in SIS to compute these values) as the total sample size increases. On the vertical axes, we give the absolute percentage relative errors of RQMC SIS and MC SIS methods. For comparison purposes, we use MC SIS as it yields the smallest error bounds in Table~\ref{tab:TLP}. The relative error bound for the MC SIS estimates are also provided. These figures suggest that RQMC SIS converges faster than MC SIS. Note that these figures are drawn for $t$-marginals and loss probability of 0.05.

\section{Conclusion}

We have combined RQMC with variance reduction techniques, IS and SIS for simulating loss probability and conditional excess under the $t$-copula model of market risk. In the implementation of RQMC, high-dimensionality and non-smoothness stand in the way as two main problems. The first problem, high-dimensionality, is solved by applying a linear transformation on the random input to reduce the effective dimension. The discontinuity of simulation integrand, the second problem, is reduced using importance sampling. We have analyzed the performance improvement of RQMC implementations of IS and SIS over their Monte Carlo implementations. The effectiveness of RQMC is illustrated in realistic portfolio examples.

\subsection*{Acknowledgments}
The authors would like to thank the Institute for Statistics and Mathematics at the Vienna University of Economics and Business for permitting the use of their computing cluster.This work was supported by Xi'an Jiaotong-Liverpool University Research Fund Project RDF-14-01-33.

\bibliographystyle{abbrvnat.bst}


\end{document}